\documentclass[%
 reprint,
 amsmath,amssymb,
 aps
]{revtex4-1}

\usepackage{graphicx}
\usepackage{dcolumn}
\usepackage{bm}
\usepackage{color}%
\usepackage{fancyhdr}
\usepackage{appendix}
\usepackage{mathrsfs}
\usepackage{enumitem}
\usepackage{footmisc}

\begin{document}
\preprint{APS/123-QED}

\title{Ultraintense Attosecond Pulse Emission from Relativistic Laser-Plasma Interaction}

\author{Suo Tang}
\email{suo.tang@mpi-hd.mpg.de}
\author{Naveen Kumar}
\affiliation{Max~Planck~Institute for Nuclear Physics, Saupfercheckweg 1, 69117 Heidelberg, Germany}%

\date{\today}

\begin{abstract}
We develop an analytical model for ultraintense attosecond pulse emission in the highly relativistic laser-plasma interaction.
In this model, the attosecond pulse is emitted by a strongly compressed electron layer around the instant when the layer transverse current changes the sign and its longitudinal velocity approaches the maximum.
The emitted attosecond pulse has a broadband exponential spectrum and a stabilized constant spectral phase $\psi(\omega)=\pm\pi/2-\psi_{A_m}$.
The waveform of the attosecond pulse is also given explicitly, to our knowledge, for the first time.
We validate the analytical model via particle-in-cell (PIC) simulations for both normal and oblique incidence.
Based on this model, we highlight the potential to generate an isolated ultraintense phase-stabilized attosecond pulse.
\end{abstract}

\pacs{Valid PACS appear here}
\maketitle

\section{Introduction}
An ultrashort pulse with atomic unit of timescale ($24$as) can be used as a camera to capture ultrafast electron dynamics in atoms, molecules and condensed matters, enabling highly time-resolved studies of many fundamental physical processes~\cite{Giuseppe:2011Npho,*Chang:2016Attosecond,Michael:2014Npho}.
Since the demonstration of attosecond pulse emission from high-order harmonic generation (HHG) in gas jets~\cite{Papadogiannis:1999PRL,*Paul:2001Sci}, remarkable developments in attosecond metrology have been utilized in extensive research areas from atomic physics to biology science~\cite{Hentschel:2001Nat,*Cavalieri:2007Nat,*Krausz:2014Npho}.
However, due to small photon flux and low photon energy of attosecond pulses generated in gaseous HHG~\cite{Ferenc:2009RMP}, the application of attosecond metrology is so far limited to low-energy physical processes in extreme ultraviolet regime.
Besides, the poor phase matching in gaseous HHG results in attosecond pulses with broad duration~\cite{Varju:2005JMO} and uncontrollable waveforms, both of which restrict the temporal resolution of attosecond metrology~\cite{Yudin:2006PRL,*Candong:2013PRL,*Marcus:2017Nphy}.
In order to extend attosecond metrology to high-energy physical processes in x-ray regime with unprecedented temporal resolution, the attosecond pulse with ultrahigh intensity, ultrabroad bandwidth and stabilized spectral phase is required, which can be achieved from HHG via ultra-relativistic laser-plasma interaction.

Plasma HHG originates from the nonlinear interaction of the electron current with a linearly polarized laser pulse at the laser-plasma interface~\cite{Teubner:2009RMP}.
Contrary to gaseous HHG in which incident laser intensities are limited to be much lower than relativistic intensity, \emph{i.e.} $I_l \ll 10^{18}~\textrm{W/cm}^2$ corresponding to $a_0=eE_{l}/m_{e}c\omega_{l} \ll 1$, due to the strong ionization of gaseous media~\cite{Ferenc:2009RMP}, much brighter harmonic photon flux can be emitted in plasma HHG by employing highly relativistic laser pulses $a_0 \gg 1$, where $I_l$, $E_{l}$ and $\omega_{l}$ are the laser intensity, electric field and frequency, $e$ and $m_{e}$ denote the electron charge and mass, $c$ is the light speed in vacuum.
As the strong laser ponderomotive force compresses the surface electrons into a layer with nanometer thickness $\Delta x \sim 1$nm~\cite{Cherednychek:2016PoP}, the collective electron current in the layer guarantees the coherence of harmonics up to $\omega\sim 1/\Delta x \sim 1$keV.
In the highly relativistic regime, plasma HHG is temporally locked around the so-called ``$\gamma$-spike" node where the plasma surface longitudinal velocity approaches the maximum~\cite{Baeva:2006PRE}, which insures the synchronization of the emission of different harmonics and results in much smaller phase chirp than that in gaseous HHG~\cite{Varju:2005JMO,Nomura:2008Nphy}.
By filtering out low-order harmonics in the phase-locking spectrum, a temporally coherent attosecond pulse train or an isolated attosecond pulse can be obtained~\cite{Tsakiris:2006NJP,Tang:2017PRE,tang2018super}.

The physics behind plasma HHG has been extensively investigated with different models~\cite{Lichters:1996PoP,Quere:2006PRL,Brugge:2010PoP,Gonoskov:2011PRE,Gonoskov:2018PoP}: coherent wake emission (CWE), relativistically oscillating mirror (ROM), coherent synchrotron emission (CSE) and relativistic electron spring (RES).
These models are associated and characterized by distinctive harmonic properties: spectrum, divergence and phase~\cite{Baeva:2006PRE,Nomura:2008Nphy,Brugge:2010PoP,Kahaly:2013PRL,Quere:2008PRL}.
The relative dominance of each model in the interaction depends on laser intensity, plasma density and interaction geometry.
Generally, the CWE model dominates in the nonrelativistic or mildly relativistic regime ($a_0\lesssim 1$) with oblique incidence on a proper plasma gradient, and in the relativistic regime ($a_0 > 1$), plasma HHG can be explained by the ROM model, while in the highly relativistic regime ($a_0 \gg 1$) the CSE and RES models are prominent with a strongly compressed electron layer at the plasma surface.
Although the emission of attosecond pulse based on these models has been predicted, the explicit waveform of the attosecond pulse, which relates straightforwardly to the applications of attosecond metrology and the plasma dynamical processes of the emission, has not been discussed in the literature so far.

In this paper, we develop a theoretical model for ultra-intense attosecond pulse emission in the ultra-relativistic laser-plasma interaction.
In Sec.\ref{Theo_model}, we first introduce the theoretical model and then derive the intensity spectrum, spectral phase and analytical waveform of the emitted attosecond pulse.
In Sec.~\ref{Simu_result}, simulation results are provided to validate the theoretical model.
At the end, a brief summary is given.
Hereafter, unless specifically stated, dimensionless quantities are used: $n_e=n_e/n_{c}$, $t=\omega_{l}t$, $x=k_{l}x$, $\beta=v/c$, $\omega_n=\omega_n/\omega_{l}$, $I=I/I_{r}$, $J=J/(ecn_c)$, $E=eE/(m_ec\omega_l)$, $B=eB/(m_e\omega_l)$, where the plasma critical density $n_c = \omega^2_l\epsilon_0m_e/e^2 = 1.742\times 10^{21}\text{cm}^{-3}$ and the relativistic laser intensity $I_{r}=c\epsilon_0(m_{e}c\omega_{l}/e)^{2}=4.276\times10^{18}\,\textrm{W/cm}^{2}$ for the laser wavelength $\lambda_l=0.8\mu$m. 

\section{Theoretical model}\label{Theo_model}

\subsection{Model description}\label{model_description}
In Fig.~\ref{Fig_comfirm}, we illustrate the evolution of the electron number density (a) and current density (b) at the plasma front surface irradiating by a highly relativistic laser pulse ($a_0=40$).
As shown, the plasma electrons are extremely compressed by the laser ponderomotive force into an ultradense layer with nanometer thickness~\cite{Cherednychek:2016PoP}.
This electron layer is crucial for the physics of laser-plasma interaction~\cite{Gonoskov:2018PoP} and dominates the plasma radiation.
The strong charge separation field formed due to the electron compression would effectively accelerate the electron layer toward the incident field to emit an intense attosecond pulse~\cite{Gonoskov:2011PRE,tang2018super}.
By tracing the pulse emission along the retardation relation $t'+x'=t+x$, where $(x,t)$ and $(x',t')$ denote the spatio-temporal points of the field detector and the plasma emitter, two characters of the pulse emission can be illuminated:

\begin{figure}
  \includegraphics[width=0.48\textwidth,height=0.39\textwidth]{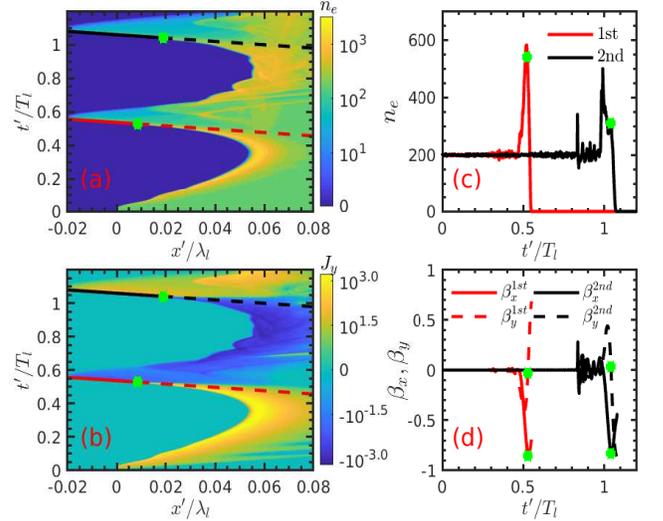}
  \caption{1D PIC simulation of the pulse emission process.
  Contour of the evolution of the electron number density $n_e$ (a) and current density $J_y$ (b) at the plasma front surface overlaid with the retardation paths of the pulse centers ($t'=x+t-x'$, red and black lines for the $1$st and $2$nd pulses in Fig.~\ref{normal_pulse} respectively.).
  Along the paths, we can trace the formation of the pulses $E_y(x,t)\propto\int J_y(x',x+t-x')dx'$.
  (c) Electron density $n_e$ along the retardation paths. 
  (d) Velocities [$\beta_x=-J_x/(en_ec)$, $\beta_y=-J_y/(en_ec)$] of the electron current along the retardation paths.
  At the emission node (green stars) where $\beta_y\approx 0$, the longitudinal velocity $\beta_x$ approaches the maximum ($\beta_x^m=-0.8518$ for the $1$st pulse and $\beta_x^m=-0.8287$ for the $2$nd pulse).
  The emission nodes in (a), (b) and (c) are also labeled by green stars.
  In (a) and (b), the solid-line parts of the retardation paths represent the pulse propagation after emissions, and the dashed-line parts denote the formation of the pulses in the plasma.
  Normal incident geometry ($\theta=0$) is employed. The laser has a step-like profile with the amplitude $a_0=40$ ,and the plasma has no pre-gradient with the constant density $n_{0}=200n_c$. Ions are free. The laser arrives the plasma surface at ($t=0$, $x=0$).}\label{Fig_comfirm}
\end{figure}

\begin{enumerate}[label=(\roman*)]
\item{\label{condition1} In Fig.~\ref{Fig_comfirm} (b), the pulse emission happens around the instant when the transverse current changes its sign, \emph{i.e.} $J_y\approx 0$ corresponding to $\beta_{y}^{el}\approx 0$.
    At this instant, the layer longitudinal velocity $\beta_{x}^{el}$ approaches its maximum in Fig.~\ref{Fig_comfirm} (d).
    We wish to stress that the condition: $\beta_{x}^{el} \approx -1$ is very important for the simplification of the derivations below, but will not affect the waveform of the attosecond pulse. The transverse current changing its sign during the emission results in the emitted pulse with an odd-functioned waveform.
}

\item{\label{condition2} In Fig.~\ref{Fig_comfirm} (c), we find that the pulse is mainly emitted by the compressed electron layer. The layer spatial distribution $f[x'-x_{el}'(t')]$ can be represented by $\delta[x'-x_{el}'(t')]$ for coherent emissions with the wavelength $\lambda_{\omega}$ much larger than the layer thickness $\Delta x$, \emph{i.e.} $\lambda_{\omega}\gg\Delta x$. We approximate the surface current for coherent emissions as
\begin{align}
J_{y}(x',t')\approx -n_{el}(t')\beta_{y}^{el}(t')\delta[x'-x_{el}'(t')]\,,
\label{theo_current}
\end{align}
where $x_{el}'(t')$, $\beta_{y}^{el}(t')$ and $n_{el}(t')$ are the location, transverse velocity and areal density of the electron layer, respectively.} 
\end{enumerate}

The radiation field from a current distribution in $1$D geometry can be expressed as
\begin{align}
 E^{r}_{y}(x,t)=-\frac{1}{2}\int^{+\infty}_{-\infty} dx'J_{y}\left(x',t'\right)\,,
\label{theo_reflection}
\end{align}
where the subscript $y$ denotes the direction of the laser electric field. Inserting Eq.~(\ref{theo_current}) into Eq.~(\ref{theo_reflection}), we obtain
\begin{align}
 E^{r}_{y}(x,t)&=\frac{1}{2}\int^{+\infty}_{-\infty} dx'n_{el}(t')\beta_{y}^{el}(t')\delta[x'-x_{el}'(t')]\nonumber\\
               &=\frac{1}{2}\int^{+\infty}_{-\infty} d\mathcal{X} \frac{n_{el}(t')\beta_{y}^{el}(t')}{1+\beta_{x}^{el}(t')}\delta(\mathcal{X})\nonumber\\
               &=\frac{n_{el}(t')[1-\beta_{x}^{el}(t')]}{2}\left.\frac{\beta_{y}^{el}(t')}{1-[\beta_{x}^{el}(t')]^2}\right|_{\mathcal{X}=0}\,,
\label{theo_step1}
\end{align}
here we replace the argument of the $\delta$-function with $\mathcal{X}=x'-x_{el}'(t')$. Because of the retardation relation $t'+x'=t+x \Rightarrow dt'=-dx'$, we can have $d\mathcal{X}=dx'-\beta_{x}^{el}(t')dt'=[1+\beta_{x}^{el}(t')]dx'$, signifying that $\mathcal{X}$ is a monotonic function of $x'$. Therefore, we can substitute the integral variable with $dx'=d\mathcal{X}/[1+\beta_{x}^{el}(t')]$. The condition $\mathcal{X}=0$ gives a new retardation relation:
\begin{align}
t+x=t'+x_{el}'(t')\,.
\label{theo_retardation}
\end{align}
Making use of the general relation $\gamma \equiv(1-\beta^2_x-\beta^2_y)^{-1/2} \Rightarrow (1-\beta^2_x)^{-1}\equiv\gamma^2[1+(\gamma\beta_y)^2]^{-1}$, we can reach
\begin{align}
 E^{r}_{y}(x,t)=\left.\frac{n_{el}(1-\beta_{x}^{el})\gamma_{el}}{4} \frac{2p_{y}^{el}}{1+(p_{y}^{el})^2}\right|_{t+x=t'+x_{el}'(t')}\,,
\label{theo_step2}
\end{align}
where $\gamma_{el}(t')$ and $p_{y}^{el}(t')=\gamma_{el}(t')\beta_{y}^{el}(t')$ are the Lorentz factor and transverse momentum of the electron layer, respectively.

Since the layer transverse current passes through the zero node during the pulse emission and the emission is on the attosecond time scale, the transverse momentum $p_{y}^{el}$ of the electron layer in the pulse emission process can be approximated in the first order:
\begin{align}
p_{y}^{el}=\Delta t'\left.\frac{dp_{y}^{el}}{dt'}\right|_{t'=t'_0}=\frac{\Delta t}{1+\beta_{x}^{el}(t'_0)}\left.\frac{dp_{y}^{el}}{dt'}\right|_{t'=t'_0}\,,
\label{theo_py}
\end{align}
here $\Delta t'$ is the short time duration around the node where $p_{y}^{el}(t'_0)=0$, and Eq.~(\ref{theo_retardation}) is utilized for the relation:
\begin{align}
\Delta t=\Delta t'[1+\beta_{x}^{el}(t'_0)]\,,
\end{align}
$\Delta t$ denotes the time duration around $t_0$ which fulfills the retardation relation $x+t_0=t'_0+x_{el}'(t'_0)$. Hereafter, we label $t_0=0$ for convenience and replace $\Delta t$ with $t$.

Inserting Eq.~(\ref{theo_py}) back into Eq.~(\ref{theo_step2}), we gain the pulse expression:
\begin{align}
E^{r}_{y}(x,t) =\hat{E^{r}_{y}} A_m \frac{2\omega_{d}t}{1+(\omega_{d}t)^2}\,,
\label{theo_step3}
\end{align}
where two crucial parameters are introduced:
\begin{align}
A_m(t')&=\left.\frac{n_{el}\gamma_{el}(1-\beta_{x}^{el})}{4}\right|_{t+x=t'+x_{el}'(t')}\nonumber\\
       &\approx \left.\frac{n_{el}\gamma_{el}}{2}\right|_{t+x=t'+x_{el}'(t')}
\label{theo_Am}
\end{align}
representing the pulse amplitude,
\begin{align}
\omega_d=\frac{1}{1+\beta_{x}^{el}(t'_0)}\left|\left.\frac{dp_{y}^{el}}{dt'}\right|_{t'=t'_0}\right|\approx2\gamma_{el}^{2}\left|\left.\frac{dp_{y}^{el}}{dt'}\right|_{t'=t'_0}\right|
\label{theo_wd}
\end{align}
scaling the pulse duration, \emph{i.e.} $T_d \sim 1/\omega_d$. We also introduce $\hat{E^{r}_{y}}=\textrm{sign}(dp_{y}^{el}/dt')$ denoting the sign of the reflected electric field, and $\beta_{x}^{el} \approx -1$ is used for simplifying Eqs.~(\ref{theo_Am}) and~(\ref{theo_wd}).

In Eq.~(\ref{theo_Am}), the pulse amplitude $A_m(t')$ depends on the retarded time. With the first order approximation: $A_m(t') = A_m(t'_0)+\Delta t'dA_m/dt'|_{t'=t'_0}$, we can transform it to depend on the real time $t$ as:
\begin{align}
A_m(t) &= A_m(t'_0)+ \frac{t}{1+\beta_{x}^{el}(t'_0)}\left.\frac{dA_m}{dt'}\right|_{t'=t'_0}\nonumber\\
       &= A^0_m+A^{1}_mt\,,
\end{align}
where $A^0_m=A_m(t'_0)$ is the constant pulse amplitude, and $A^{1}_m=dA_m/dt'/[1+\beta_{x}^{el}(t')]|_{t'=t'_0}$ denotes the first order temporal derivative of the pulse amplitude.
The straightforward consequence of this amplitude variation is pulse asymmetry and a constant shift of the pulse spectral phase as we will see later.

As we see, the pulse amplitude $A_m$ depends on the product of the areal density $n_{el}$ and the relativistic Lorentz factor $\gamma_{el}$ of the electron layer, and $\omega_d$ is determined by the layer transverse acceleration $|dp_{y}^{el}/dt'|$ and also the Lorentz factor $\gamma_{el}$. These may point out the direction to generate a more intense pulse with shorter duration by tailoring the laser-plasma parameters to increase the value of $n_{el}$,$\gamma_{el}$ and $|dp_y^{el}/dt'|$.

The above derivations are based on the dynamic properties of the electron layer and do not take advantage of any specific effects, \emph{e.g.} hole-boring effect, collision damping, temperature effect, or radiation reaction force etc. All of these effects can be taken into account for the pulse emission by considering their influence on the kinetic parameters ($n_{el}$,$\gamma_{el}$, $|dp_{y}^{el}/dt'|$) of the electron layer. For example, based on the plasma equations of motion~\cite{Kruer_book}:
\begin{subequations}
\label{motion_1d}
 \begin{align}
\frac{d}{d t}p_{x}^{el}&=-(E_x+\beta_{y}^{el}B_z)-\frac{1}{n_e}\frac{\partial}{\partial x}\tilde{\tilde{P}}_{xx}-\nu_{c}p_{x}^{el}\,,\label{motion_1d_a}\\
\frac{d}{d t}p_{y}^{el}&=-(E_y-\beta_{x}^{el}B_z)-\frac{1}{n_e}\frac{\partial}{\partial x}\tilde{\tilde{P}}_{xy}-\nu_{c}p_{y}^{el}\,,\label{motion_1d_b}
\end{align}
\end{subequations}
where
\begin{subequations}
 \begin{align}
\tilde{\tilde{P}}_{xx}=\int(\beta_x-\beta_{x}^{el})(p_x-p_{x}^{el})f_e d\textbf{p}\,,\nonumber\\
\tilde{\tilde{P}}_{xy}=\int(\beta_x-\beta_{x}^{el})(p_y-p_{y}^{el})f_e d\textbf{p}\,,\nonumber
\end{align}
\end{subequations}
are the elements of the plasma thermal pressure tensor $\tilde{\tilde{P}}=\int(\bm{\beta}-\bm{\beta^{el}})(\textbf{p}-\bm{p^{el}})f_e d\bm{p}$, $\nu_{c}$ is the collision frequency, $f_e(\bm{p},x,t)$ is the electron distribution function in the phase space ($x$, $p_x$, $p_y$) and $\bm{\beta}=\bm{p}/\sqrt{p^2+1}$, we can clearly see that collision effect damps the motion of the electron layer, and that the temperature pressure impedes the longitudinal compression and transverse acceleration of the electron layer. Both of the effects lead to smaller pulse amplitude $A_m$ and longer duration $1/\omega_d$.

In the cold fluid approximation, we have $\tilde{\tilde{P}}=0$, and ignore the collision damping. The temporal derivative of the transverse momentum can be approximated as
\begin{align}
\left.\frac{dp_{y}^{el}}{dt'}\right|_{t'=t'_0} &= \left.-(E_y-\beta_{x}^{el}B_z)\right|_{t'=t'_0}\nonumber\\
                 &=\left.-\left[(E^{i}_y-\beta_{x}^{el}B^{i}_z)+(E^{r}_y-\beta_{x}^{el}B^{r}_z)\right]\right|_{t'=t'_0}\nonumber\\
                 &\approx -\hat{E^{i}_{y}}2|E^{i}_y(t'_0)|\,,
\label{theo_py_derivative}
\end{align}
where $\hat{E^{i}_{y}}$ is the sign of the incident electric field at the emission instant $t'_0$, and $E^{r}_y=-B^{r}_z$ is used because of the reflected pulse propagating in $-x$ direction. Inserting Eq.~(\ref{theo_py_derivative}) into Eq.~(\ref{theo_wd}), we can have
\begin{align}
\omega_d\approx4\gamma_{el}^{2}(t'_0)\left|E^{i}_y(t'_0)\right|\,,
\label{theo_wd_cold}
\end{align}
and relate the sign of the reflected field to the sign of the incident field, \emph{i.e.} $\hat{E^{r}_{y}}=-\hat{E^{i}_{y}}$.

\subsection{Spectral and phase properties}\label{Theo_spectral_Phase}
The spectral and phase properties relate closely to the applications of the emitted pulse in experiments.
A pulse with an ultrabroad spectrum is always needed for inner shell electron excitation in high-Z atoms~\cite{Michael:2014Npho} and a well-locked spectral phase is crucial for the coherent control of the excitation processes~\cite{Prakelt:2004PRA,*Ryoichi:2017PRL}.

From Eq.~(\ref{theo_step3}), the pulse spectrum can be calculated via a simple Fourier transformation:
\begin{align}
\widetilde{E}^{r}_{y}(\omega) &= \frac{-\hat{E^{i}_{y}}}{2\pi}\int^{\infty}_{-\infty}A_m(t) \frac{2\omega_{d}t}{1+(\omega_{d}t)^2}e^{i\omega t} dt\nonumber\\
             &= \frac{-\hat{E^{i}_{y}}}{2\pi\omega_d}\int^{\infty}_{-\infty}\left(A^0_m+\frac{A^{1}_m}{\omega_d}\mathcal{X}\right) \frac{2\mathcal{X}}{1+\mathcal{X}^2}e^{i\frac{\omega}{\omega_d} \mathcal{X}} d\mathcal{X}\nonumber\,,
\end{align}
where the integral variable is replaced by $\mathcal{X}=\omega_d t \Rightarrow dt=d\mathcal{X}/\omega_d$. 
If $\omega>0$,
\begin{align}
\widetilde{E}^{r}_{y}(\omega) &= \frac{-\hat{E^{i}_{y}}}{2\pi\omega_d}\int^{\infty}_{-\infty}\left(A^0_m+\frac{A^{1}_m}{\omega_d}\mathcal{X}\right) \frac{1}{\mathcal{X}-i} e^{i\frac{\omega}{\omega_d} \mathcal{X}} d\mathcal{X}\nonumber\\
             &= \frac{-\hat{E^{i}_{y}}}{2\pi\omega_d} 2\pi i\left(A^0_m+\frac{A^{1}_m}{\omega_d}i\right)e^{-\frac{\omega}{\omega_d}}\nonumber\\
             &=\frac{\bar{A}_m}{\omega_d} e^{-\frac{\omega}{\omega_d}}e^{-i(\hat{E^{i}_{y}}\frac{\pi}{2}-\psi_{A_m})}\,,
\label{theo_phase_residue_A}
\end{align}
if $\omega<0$,
\begin{align}
\widetilde{E}^{r}_{y}(\omega) &= \frac{-\hat{E^{i}_{y}}}{2\pi\omega_d}\int^{\infty}_{-\infty}\left(A^0_m+\frac{A^{1}_m}{\omega_d}\mathcal{X}\right) \frac{1}{\mathcal{X}+i} e^{i\frac{\omega}{\omega_d} \mathcal{X}} d\mathcal{X}\nonumber\\
             &= \frac{-\hat{E^{i}_{y}}}{2\pi\omega_d} (-2\pi i)\left(A^0_m-\frac{A^{1}_m}{\omega_d}i\right)e^{\frac{\omega}{\omega_d}}\nonumber\\
             &=\frac{\bar{A}_m}{\omega_d} e^{\frac{\omega}{\omega_d}}e^{-i(-\hat{E^{i}_{y}}\frac{\pi}{2}+\psi_{A_m})}\,,
\label{theo_phase_residue_B}
\end{align}
and if $\omega=0$,
\begin{align}
\widetilde{E}^{r}_{y}(0) =-\hat{E^{i}_{y}}\frac{A_m^{1}}{\omega^2_d}\left[2\omega_d\left.\delta(\omega)\right|_{\omega=0}-1\right]\,,
\end{align}
where
\begin{subequations}
\begin{align}
\bar{A}_m &= \sqrt{(A^0_m)^2+\left(A^{1}_m/\omega_d\right)^2}\,,\\
\cos(\psi_{A_m}) &= \frac{A^0_m\omega_d}{\sqrt{(A^0_m\omega_d)^2+(A^{1}_{m})^2}}\,,\\
\sin(\psi_{A_m}) &= \frac{A^{1}_{m}}{\sqrt{(A^0_m\omega_d)^2+(A^{1}_{m})^2}}\,.
\end{align}
\label{theo_definition_AM_phi}
\end{subequations}

From this derivation, we can obtain the pulse spectrum:
\begin{align}
I(\omega)=|\widetilde{E}^{r}_{y}(\omega)|^2=\frac{\bar{A}^{2}_{m}}{\omega^{2}_{d}}e^{-2\frac{|\omega|}{\omega_d}}\,.
\label{theo_spectrum}
\end{align}
The emitted pulse possesses an exponential spectrum with the spectral decay $2/\omega_d$.
As the Lorentz factor $\gamma_{el}$ and the transverse acceleration $|dp_{y}^{el}/dt'|$ of the electron layer are boosted in the ultra-relativistic regime, the spectral decay $2/\omega_d \propto \gamma_{el}^{-2}|dp_{y}^{el}/dt'|^{-1}$ would become very slow, implying an ultrabroadband pulse with ultrashort duration $T_d\sim 1/\omega_d$.

Furthermore, we can also obtain the pulse spectral phase $\psi(\omega)$:
\begin{subequations}
\begin{align}
\psi(\omega)&=~~\hat{E^{i}_{y}}\frac{\pi}{2}-\psi_{A_m}\,,~~~~~\textrm{for}~~~\omega>0\,,\\
\psi(\omega)&=-\hat{E^{i}_{y}}\frac{\pi}{2}+\psi_{A_m}\,,~~~~~\textrm{for}~~~\omega<0\,.
\end{align}
\label{Spectral_Phase_psi}
\end{subequations}
with the definition~\footnote{\label{phase_sign}The sign of $\psi(\omega)$ is chosen to be same with the linear term $\omega t$ in $E(t)=\int^{\infty}_{-\infty}|E(\omega)|e^{-i[\omega t+\psi(\omega)]}dt$.}: $\widetilde{E}^{r}_{y}(\omega)=\left|\widetilde{E}^{r}_{y}(\omega)\right|\exp(-i\psi(\omega))$
As we see, the pulse spectral phase is a constant and composed of two terms:
\begin{enumerate}
\item{$\pi/2$: This particular phase is the consequence of the transverse current changing its sign at the emission instant when $\beta_y^{el}=0$.
This term regulates the pulse waveform and leads to a minimum at the pulse center, contrary to the synchrotron-like pulse~\cite{Mikhailova:2012PRL}.
We wish to stress that this particular phase does not depend on the carrier-envelope-phase (CEP) of the incident laser, but on the dynamics of the compressed electron layer during the emission.
This property stabilizes the spectral phase of the generated attosecond pulses from the laser pulses with shot-to-shot unstabilized CEP.}

\item{$\psi_{A_m}$: This term arises from the temporal variation of the pulse amplitude $A_m$ and would induce the pulse waveform asymmetry.
  In principle, this phase depends on the CEP of the incident laser because the pulse amplitude $A_m(t)$ relates to the processes of electron layer compression ($n_{el}$) and acceleration ($\gamma_{el}$), both can change in the interaction driven by the laser with different CEP.
  However, in the ultra-relativistic regime, the dependence is negligible as the phase $\psi_{A_m}$ itself would become very small.
  In this regime, the duration ($\propto 1/\omega_d$) of pulse emission is extremely short, the value of the temporal variation $A^1_m/\omega_d$ would be much smaller than the constant value $A^0_m$, \emph{i.e.} $A^1_m/\omega_d \ll A^0_m$, and thus $\psi_{A_m}$ can be approximated as
    \begin{align}
      \psi_{A_m}\sim A^1_{m}/(A^0_m\omega_d)\,.
      \label{theo_Spectral_Phase_simp_psi_AM}
    \end{align}
    }
\end{enumerate}
We point out that the pulse spectral phase denotes the \emph{time-independent} phase of the different frequency components in a single pulse, it relates directly to the dynamics of the surface layer during the pulse emission. By studying the spectral phase of each pulse, we can diagnose the plasma dynamics on attosecond time scale which may be hidden in the harmonic phase. The latter is the consequence of the interference among all the pulses in the whole reflection~\cite{Varju:2005JMO,Quere:2008PRL,Nomura:2008Nphy}.

\subsection{Finite distribution of the electron layer}\label{theo_atto_Finite}
If the emission wavelength $\lambda_{\omega}$ is close to or smaller than the layer thickness $\lambda_{\omega} \lesssim \Delta x$, the $\delta$-function approximation of the layer distribution [see Eq.~(\ref{theo_current})] can not be applied.
The pulse spectrum and the spectral phase in this high-frequency region must be modulated.

We now calculate the spectrum from Eq.~(\ref{theo_reflection}) with the finite extension of the current density
\begin{align}
J_y(t',x')=-n_{el}(t')\beta_{y}(t')f[x'-x_{el}'(t')]\,,\nonumber
\end{align}
where we neglect the temporal dependence of the finite distribution: $f(x',t') \approx f(x')$. Because of the extremely short emission duration, the electron layer expansion is negligible.
Thus we can have
\begin{align}
 \widetilde{E}^{r}_{y}(x,\omega) &=\frac{-1}{4\pi}\left.\int^{\infty}_{-\infty} dt e^{i\omega t}\int^{\infty}_{-\infty} dx'J_{y}(x',t')\right|_{x+t=x'+t'} \nonumber\\
                                &=e^{-i\omega x}\widetilde{F}(-\omega)\cdot\textrm{Const}\,,
\label{theo_finite_step1}
\end{align}
where the Fourier expansion of the finite spatial distribution is used: $f(x')=\int^{\infty}_{-\infty}\widetilde{F}(k)e^{ikx'}dk$, and \textrm{Const} is an integral constant:
\begin{align}
 \textrm{Const}=\int^{\infty}_{-\infty} \frac{n_{el}(t')\beta_{y}(t')}{2} e^{i\omega [t'+x_{el}'(t')]} dt'\nonumber\,.
\end{align}

The integral constant can be evaluated by the substitution: $\mathcal{X}=t'+x_{el}'(t') \Rightarrow dt'=d\mathcal{X}/[1+\beta_x(t')]$:
\begin{align}
 \textrm{Const} &= \int^{\infty}_{-\infty} \left.A_m(t')\frac{2p_y(t')}{1+p^2_y(t')} e^{i\omega \mathcal{X}} d\mathcal{X}\right|_{\mathcal{X}=t'+x_{el}'(t')}\nonumber\\
                                 &\approx-\hat{E^{i}_{y}}\int^{\infty}_{-\infty} \left.A_m(\mathcal{X}-\mathcal{X}_0)\frac{2\omega_d (\mathcal{X}-\mathcal{X}_0)}{1+\omega^2_d(\mathcal{X}-\mathcal{X}_0)^2} e^{i\omega \mathcal{X}} d\mathcal{X}\right.
                                 \nonumber\\
                                 &=-e^{i\omega\mathcal{X}_0}\hat{E^{i}_{y}}\int^{\infty}_{-\infty} A_m(t)\frac{2\omega_d t}{1+\omega^2_dt^2} e^{i\omega t} dt\,,
\label{theo_finite_step3}
\end{align}
where $\mathcal{X}_0=t'_0+x_{el}'(t'_0)$.
We consider the main contribution around the emission instant when $\beta_y(t'_0)=0$, $\beta_x(t'_0)\approx -1$. This is in line with the stationary phase approximation in Ref.~\cite{Brugge:2010PoP}.
During the emission, the phase term $\exp(i\omega \mathcal{X})$ is approximated to be constant because of $d\mathcal{X}=dt'(1+\beta_x)\approx0$, which mainly contributes to the integral.
At the non-stationary phase point, the phase term results in rapid oscillations in the integral, especially for high frequencies, thus their contributions cancel each other and can be neglected~\cite{wong2001asymptotic}.

After the same calculations in Eqs.~(\ref{theo_phase_residue_A}) and~(\ref{theo_phase_residue_B}), we obtain
\begin{align}
 \widetilde{E}^{r}_{y}(x,\omega) = & 2\pi\left|\widetilde{F}(\omega)\right|\frac{\bar{A}_m}{\omega_d}e^{-\frac{|\omega|}{\omega_d}}\nonumber\\
                                   &e^{-i\psi_f(\omega)}
 \begin{cases}
    e^{-i(~~\hat{E^{i}_{y}}\frac{\pi}{2}-\psi_{A_m})}, & \omega>0,\\
    e^{-i(-\hat{E^{i}_{y}}\frac{\pi}{2}+\psi_{A_m})},&  \omega<0,\\
 \end{cases}
\label{theo_finite_step4}
\end{align}
and
\begin{align}
 I(\omega)=|\widetilde{E}^{r}_{y}(x,\omega)|^{2}=4\pi^2\frac{|\bar{A}_{m}|^2}{\omega_d^2} |\tilde{F}(\omega)|^2 e^{-2\frac{\omega}{\omega_d}}\,,
\label{theo_finite_spectrum}
\end{align}
where $\widetilde{F}(-\omega)=\widetilde{F}^{*}(\omega)=|\widetilde{F}(\omega)|e^{-i\psi_f(\omega)}$ is used, and $t_0=t'_0+x_{el}'(t'_0)-x$ is taken to be zero. From Eqs.~(\ref{theo_finite_step4}) and~(\ref{theo_finite_spectrum}), we see that the layer finite distribution can affect not only the pulse spectrum with $|\tilde{F}(\omega)|^2$ but also the spectral phase with $\psi_f(\omega)$.

To clearly see the influence of the layer finite distribution, we qualitatively express $\widetilde{F}(\omega)$ as
\begin{align}
\widetilde{F}(\omega) &=\frac{1}{2\pi}\int^{\infty}_{-\infty} f(x) e^{-i\frac{\omega}{c} x} dx\nonumber\\
                      &\approx\frac{1}{2\pi}\int^{x_{el}+\frac{\Delta x}{2}}_{x_{el}-\frac{\Delta x}{2}} f(x) e^{-i\frac{2\pi x}{\lambda_{\omega}}}dx\,.\nonumber
\end{align}
If the wavelength $\lambda_{\omega} \gg \Delta x$, $2\pi x/\lambda_{\omega}$ approaches to zero, we have
\begin{align}
\widetilde{F}(\omega) \approx\frac{1}{2\pi}\int^{x_{el}+\Delta x/2}_{x_{el}-\Delta x/2} f(x)dx = \frac{1}{2\pi}\nonumber
\end{align}
which means that for low-frequency emissions ($\lambda_{\omega} \gg \Delta x$), the layer finite distribution does not affect the pulse spectrum or the spectral phase.

If $\lambda_{\omega} \lesssim \Delta x$, the phase $2\pi x/\lambda_{\omega}$ would be significant and result in rapid oscillation in the integral, thus
\begin{align}
\widetilde{F}(\omega) \approx\frac{1}{2\pi}\int^{x_{el}+\Delta x/2}_{x_{el}-\Delta x/2} f(x) e^{-i2\pi\frac{x}{\lambda_{\omega}}}dx \ll 1\nonumber
\end{align}
which indicates that the layer finite distribution could speed up the decay of the high-frequency spectrum ($\lambda_{\omega} \lesssim \Delta x$), and simultaneously, the distribution-induced phase $\psi_f(\omega)$ would also be very important for the spectral phase of the high-frequencies.

On the other hand, we assume that different parts of the electron layer contribute to the pulse emission coherently.
Since the radiations from different parts may arrive at detector at different times, the time difference $\Delta t \propto \Delta x/c$ leads to a phase difference $\Delta \psi(\omega)\approx \omega\Delta x/c$, thus affecting the intensity of the pulse which actually is the superposition of the radiations from all parts of the electron layer.
Obviously, this incoherence is negligible for radiations with $\lambda_{\omega}=2\pi c/\omega \gg \Delta x$, but could induce significant phase fluctuation for emissions with $\lambda_{\omega} \lesssim \Delta x$.
Hence, based on the layer thickness, we can qualitatively define a threshold for incoherent emission:
\begin{align}
\omega_{in}^{th}\approx \frac{2\pi c}{\Delta x}\,,
\label{threshold}
\end{align}
which actually truncates the region of coherent emission.
Moreover, the phase space $f(\beta_x,\beta_y)$ of the electron layer is expanded, implying that only electrons in a narrow phase space element emit coherently.

\subsection{Attosecond pulse generation}\label{theo_atto_pulse}
With the above discussions, we know that the coherent high-frequency emissions are bunched on attosecond time scale around the node where $\beta_y=0$, $\beta_x\approx -1$.
By filtering out the low-frequencies in the reflection, an attosecond pulse can be manifested.

To obtain the analytical waveform of the attosecond pulse, we filter out the low-frequency components ($\omega<\omega_f$) in Eqs.~(\ref{theo_phase_residue_A}) and~(\ref{theo_phase_residue_B}), and then inversely transform the coherent high-frequency components back in the time domain:
\begin{align}
E^{r}_{y}(\omega_f,t) =& \int^{\omega_{in}^{th}}_{\omega_f}\widetilde{E}^{r}_{y}(\omega)e^{-i\omega t} d\omega+\int^{-\omega_f}_{-\omega_{in}^{th}}\widetilde{E}^{r}_{y}(\omega)e^{-i\omega t} d\omega\nonumber\\
                        =&-\hat{E^{i}_{y}}\frac{2\bar{A}_m}{\omega_d}\cos(\psi_{A_m})\int^{\omega_{in}^{th}}_{\omega_f} e^{-\frac{\omega}{\omega_d}} \sin(\omega t) d\omega \nonumber\\
                        &+\hat{E^{i}_{y}}\frac{2\bar{A}_m}{\omega_d}\sin(\psi_{A_m})\int^{\omega_{in}^{th}}_{\omega_f} e^{-\frac{\omega}{\omega_d}} \cos(\omega t) d\omega\nonumber\,.
\end{align}

Here, we introduce two integral constants:
\begin{align}
  C_s(\omega_a,\omega_d) &=~\int^{+\infty}_{\omega_a} e^{-\frac{\omega}{\omega_d}} \sin\left(\omega t\right) d\omega\,, \nonumber\\
  C_c(\omega_a,\omega_d) &=~\int^{+\infty}_{\omega_a} e^{-\frac{\omega}{\omega_d}} \cos\left(\omega t\right) d\omega\,, \nonumber
\end{align}
after simple calculations, we can gain:
\begin{align}
  C_s(\omega_a,\omega_d) &=\frac{\omega_d}{1+\omega_d^2t^2} e^{-\frac{\omega_a}{\omega_d}}\left[\sin(\omega_a t)+\omega_dt\cos(\omega_at)\right]\,, \nonumber\\
  C_c(\omega_a,\omega_d) &=\frac{\omega_d}{1+\omega_d^2t^2} e^{-\frac{\omega_a}{\omega_d}}\left[\cos(\omega_a t)-\omega_dt\sin(\omega_at)\right]\,. \nonumber
\end{align}

Making use of these two integral constants, we can obtain an explicit expression for the attosecond pulse:
\begin{align}
E^{r}_{y}(\omega_f,t) =&-\hat{E^{i}_{y}}\frac{2\bar{A}_m}{\omega_d}\nonumber\\
                       &\left\{\cos(\psi_{A_m})[C_s(\omega_f,\omega_d)-C_s(\omega_{in}^{th},\omega_d)]\right.\nonumber\\
                      -&\left.\sin(\psi_{A_m})[C_c(\omega_f,\omega_d)-C_c(\omega_{in}^{th},\omega_d)]\right\}\nonumber\\
                      =&\frac{-2\hat{E^{i}_{y}}\bar{A}_m}{\sqrt{1+(\omega_dt)^2}}\left\{e^{-\frac{\omega_f}{\omega_d}}\cos\left[\omega_ft+\varphi(t)-\psi_{A_m}\right]\right.\nonumber\\
                      &~~~~-\left.e^{-\frac{\omega_{in}^{th}}{\omega_d}}\cos\left[\omega_{in}^{th}t+\varphi(t)-\psi_{A_m}\right]\right\}
\label{theo_AS_T}
\end{align}
with the definition of the temporal phase chirp $\varphi(t)$:
\begin{align}
\cos[\varphi(t)]=\frac{\omega_dt}{\sqrt{1+(\omega_dt)^2}},~~\sin[\varphi(t)]=\frac{-1}{\sqrt{1+(\omega_dt)^2}}\nonumber\,.
\end{align}

In order to obtain a strong attosecond pulse, the filtering frequency must be much smaller than the threshold of incoherent emission, \emph{i.e.} $\omega_f \ll \omega_{in}^{th}$ $[\exp(-\omega_f/\omega_d) \gg \exp(-\omega_{in}^{th}/\omega_d)]$, thus we can write the attosecond pulse expression in a simplified form:
\begin{align}
  E^{r}_{y}(\omega_f,t)=\frac{-2\hat{E^{i}_{y}}\bar{A}_m e^{-\frac{\omega_f}{\omega_d}}}{\sqrt{1+(\omega_dt)^2}}\cos\left[\omega_f t+\varphi(t)-\psi_{A_m}\right]\,.
\label{theo_AS_compact}
\end{align}
As we can see, the amplitude of the attosecond pulse:
\begin{align}
  A_{atto} = 2\bar{A}_m e^{-\frac{\omega_f}{\omega_d}}
\label{theo_AS_Amplitude}
\end{align}
depends linearly on the original pulse amplitude $\bar{A}_m$ but exponentially on the ratio of the filtering frequency $\omega_f$ to the parameter $\omega_d$. The pulse temporal envelope:
\begin{align}
  f_{atto}(t) = \frac{1}{\sqrt{1+(\omega_dt)^2}}
\label{AS_profile}
\end{align}
is determined only by $\omega_d$. The attosecond pulse duration can be calculated as the full width at half maximum (FWHM) of the intensity profile $[f^2_{atto}(t)]$ and is given as
\begin{align}
 T_d=\frac{2}{\omega_d}\,.
\label{theo_AS_duration}
\end{align}
From Eq.~(\ref{theo_AS_compact}), we can also know that the oscillation of the electric field is controlled by $\omega_ft$ and the temporal chirp $\varphi(t)$. If we set the filtering frequency $\omega_f \lesssim \omega_d$, the electric field oscillation mainly depends on $\varphi(t)$, leading to the generation of an ultrashort single-cycle attosecond pulse. The constant phase $\psi_{A_m}$ can be regarded as the CEP of the attosecond pulse.

In the ultra-relativistic regime, $\omega_d$ becomes very large, leading to $\psi_{A_m}\thickapprox 0$ [see Eq.~(\ref{theo_Spectral_Phase_simp_psi_AM})], then the attosecond pulse expression can be further simplified as~\cite{tang2018super}
\begin{align}
  E^{r}_{y}(\omega_f,t)=\frac{-2\hat{E^{i}_{y}}A_m}{\sqrt{1+(\omega_dt)^2}}e^{-\frac{\omega_f}{\omega_d}}\cos[\omega_{f}t+\varphi(t)]\,,
\label{theo_AS_simplify}
\end{align}
which demonstrates the probability to generate an ultraintense, ultrabroadband and phase-stabilized attosecond pulse.

\subsection{Comparison with the RES and CSE models}
The exponential spectrum is same as in the RES model~\cite{Gonoskov:2011PRE}, but expressed in a different parametric form.
This is apparent since the RES model considers the radiation from an ideal moving electron layer.
However, as it ignores the temporal variation of the pulse amplitude during the emission, \emph{i.e.} $A^{1}_{m}=0$, the RES model cannot explain the phase property [see Eq.~(\ref{Spectral_Phase_psi})] of the emitted pulse.
Moreover, the condition $\beta^2_x+\beta^2_y=1$ utilized in the RES model, in order to self-consistently evolve the motion equations, results in a singularity in Eq.~(\ref{theo_step3}) at the pulse emission node where $\beta_y=0$. Thus it cannot describe the realistic waveform of the attosecond pulse.

In our model, the low-frequency emission during the layer backward acceleration is neglected, as we intend to study the properties of attosecond pulse which is synthesized by the emitted high-frequencies when $\beta_x$ approaches its maximum.
Though the consideration of the layer acceleration, $\beta_{x}(t')=-\beta_{0}+\alpha t'^2$ in the CSE model~\cite{Brugge:2010PoP}, gives a power-law spectrum [$I(\omega) \propto \omega^{-4/3}$] in low-frequency region, the emitted attosecond pulse cannot be changed as the low-frequencies have to be filtered off. Moreover, the analytical calculation of the pulse waveform becomes very challenging with the consideration of the layer acceleration because of the Airy function involved as discussed in the CSE model~\cite{Brugge:2010PoP}.

\section{Simulation results}\label{Simu_result}
To confirm our analytical model, we consider a fully ionized carbon plasma ($Z/A=6/12$) irradiated respectively by a highly relativistic laser pulse ($a_0=40$) with normal ($\theta=0$) incidence and by an ultra-relativistic laser pulse ($a_0=100$) with oblique ($\theta=45^{\circ}$) incidence, where $Z$ and $A$ denote the charge and mass number of the carbon ion, $\theta$ is the laser incident angle.
Both of the cases are simulated in $1$D geometry with the EPOCH-PIC code~\cite{Arber:2015PPCF} including the effect of plasma collisions.

\subsection{Normal incidence}\label{normal}
\begin{figure}
  \includegraphics[width=0.45\textwidth,height=0.32\textwidth]{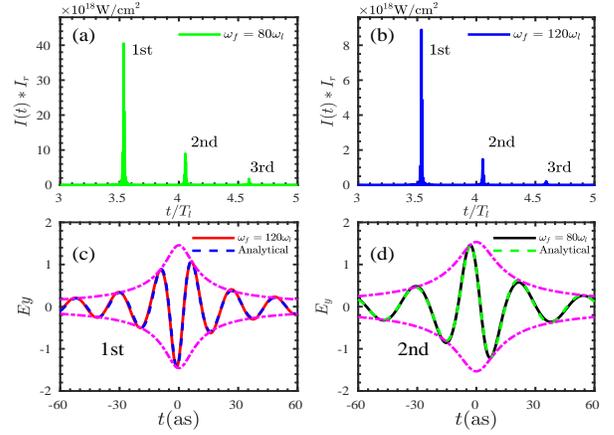}
 \caption{Attosecond pulses obtained by applying different filter frequency $\omega_f$: (a) $\omega_f=80\omega_l$, (b) $\omega_f=120\omega_l$.
 (c), (d) Electric field of the $1$st pulse with $\omega_f=120\omega_l$ and the $2$nd pulse with $\omega_f=80\omega_l$ compared with the analytical results, respectively. The temporal envelope [see Eq.~(\ref{AS_profile}), magenta dashed-dotted line] of each pulse is also shown.
 We label the pulse center at time $t=0$ and zoom in the time axis in unit of attosecond (as).
 The upper frequency of the filter is $500 \omega_l$.
 The field detector is located at $3\lambda_l$ from the plasma surface.}\label{normal_pulse}
\end{figure}

In Fig.~\ref{normal_pulse}, we show the obtained attosecond pulses by filtering out low-order harmonics ($\omega<\omega_f$) in the reflection from the laser-plasma parameters described in Fig.~\ref{Fig_comfirm}.
The retardation paths of the $1$st and $2$nd pulses are shown in Fig.~\ref{Fig_comfirm} (a) and (b).
As clearly shown, in each half laser cycle, one attosecond pulse is emitted by the backward accelerated electron layer around the instant when the transverse current changing sign.

\begin{figure}
  \includegraphics[width=0.45\textwidth]{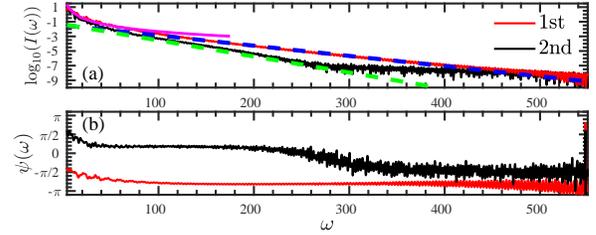}
  \caption{(a) Intensity spectra and (b) spectral phase of the $1$st and $2$nd pulses from Fig.~\ref{normal_pulse}. The dashed lines in (a) are the spectral fitting for the $1$st [$\textrm{log}_{10}(I)=-1.47-0.014\omega$, blue] and $2$nd [$\textrm{log}_{10}(I)=-1.15-0.022\omega$, green] pulses. The magenta solid line is the power-law fitting spectra $[I(\omega)\propto \omega^{-8/3}]$ for the lower frequency region.}\label{normal_spectral_phase}
\end{figure}

\begin{figure}[t]
  \includegraphics[width=0.45\textwidth]{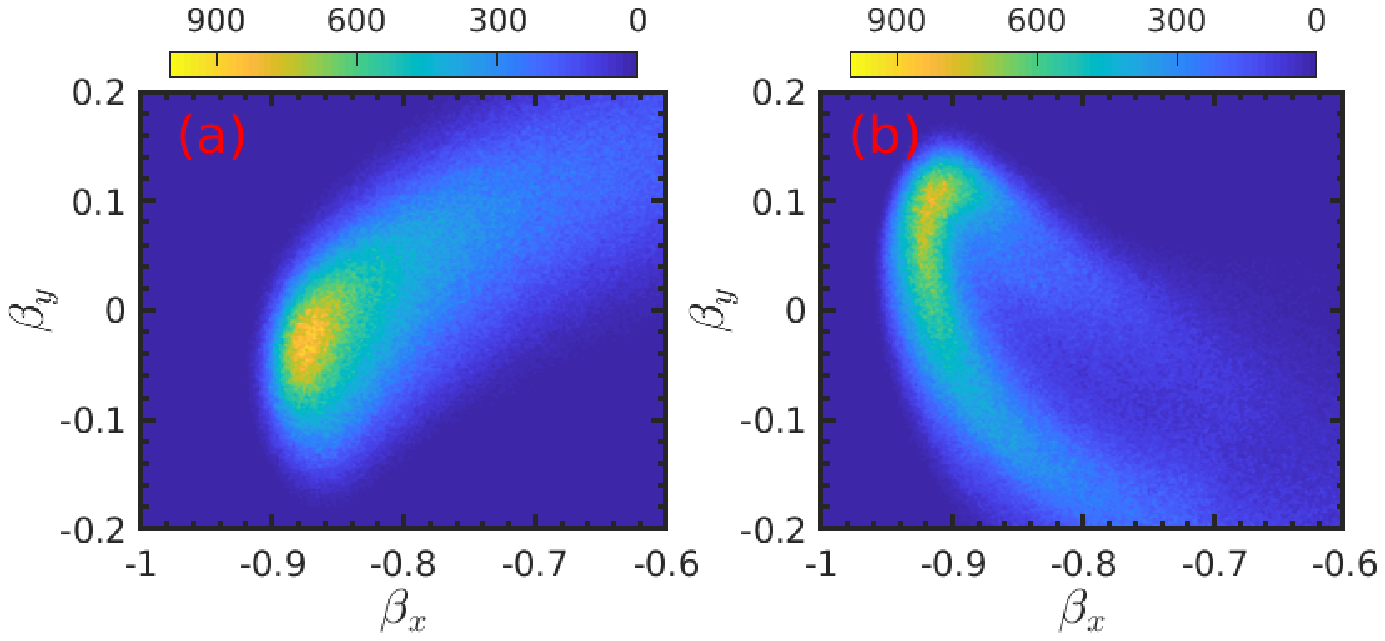}\\
  \includegraphics[width=0.45\textwidth]{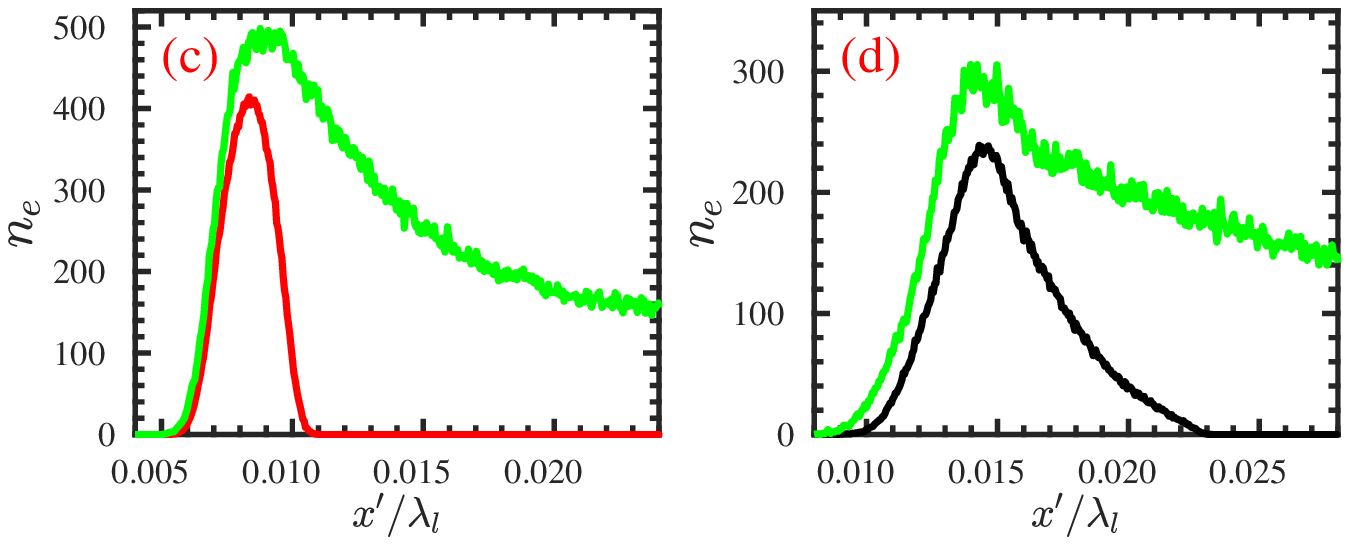}\\
  \includegraphics[width=0.45\textwidth]{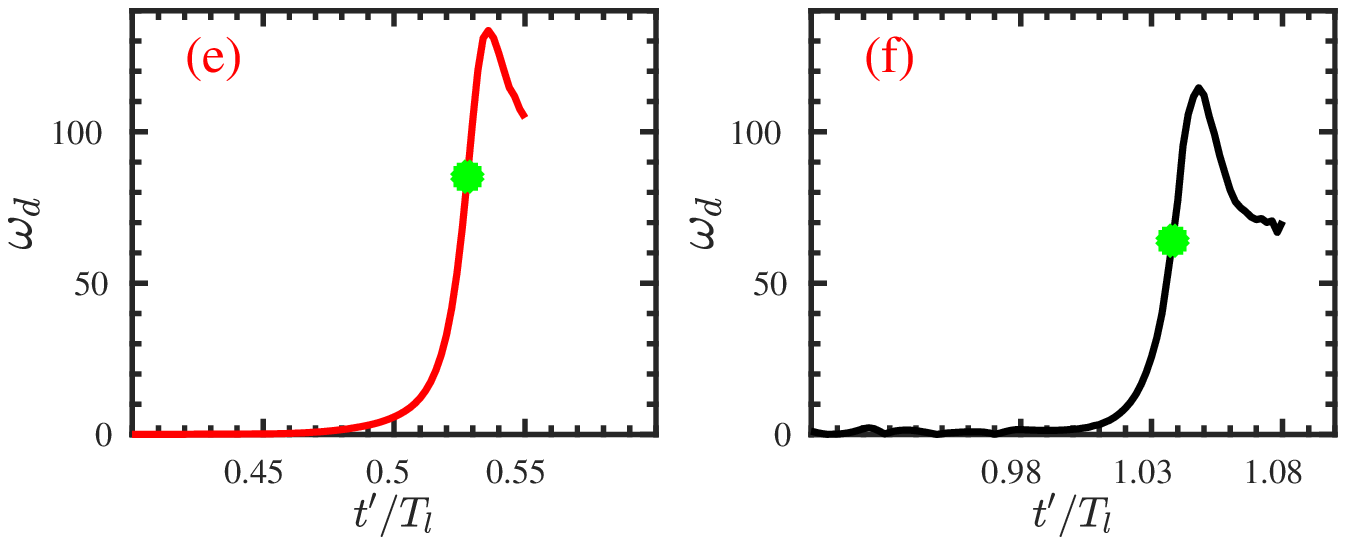}
  \caption{(a), (b) Phase space distributions $f(\beta_x,\beta_y)$ of the electron layer at the emission instants;
  (c), (d) Density distributions $n_e(x)$ of the plasma electrons at the emission instants (green line). The density of electrons in the phase space element: $\beta_x<-0.8$ and $|\beta_y|<0.1$ are also shown (red and black lines);
  (e), (f) Evaluation of the parameter $\omega_d$ along the retardation paths as discussed in Fig.~\ref{Fig_comfirm}. Green stars denote the emission instants;
  (a), (c) and (e) are for the $1$st pulse; (b), (d) and (f) are for the $2$nd pulse.}\label{normal_pulse_wd}
\end{figure}

In Fig.~\ref{normal_spectral_phase}, we present the intensity spectra and the spectral phase of the $1$st and $2$nd pulses from Fig.~\ref{normal_pulse}.
As we can see in Fig.~\ref{normal_spectral_phase} (a), the emitted pulses have broad exponential spectra in the regions: $80<\omega<450$ for the $1$st pulse, and $50<\omega<280$ for the $2$nd pulse.
The exponential regions are numerically confirmed by the linear-logarithm fitting [blue and green dashed lines in Fig.~\ref{normal_spectral_phase} (a)]: $\log_{10}[I(\omega)]=\log_{10}(I_{0})-S_{k}\omega$.
Comparing with Eq.~(\ref{theo_spectrum}) and utilizing the spectral fitting slope $S_{k}$, we can obtain the spectral decay parameter $\omega_d$ of each pulse:
\begin{subequations}
\label{simu_pulse_wd}
\begin{align}
\frac{2\log_{10}(e)}{\omega_d}&=0.014\Rightarrow\omega_d=62.04,~\textrm{for}~1\textrm{st pulse}\,,\\
\frac{2\log_{10}(e)}{\omega_d}&=0.022\Rightarrow\omega_d=39.48,~\textrm{for}~2\textrm{nd pulse}\,.
\end{align}
\end{subequations}
Inserting $\omega_d$ into Eq.~(\ref{AS_profile}), we can exactly produce the envelopes of attosecond pulses in Fig.~\ref{normal_pulse} (c) and (d), which give the pulse duration: $T_d=2/\omega_d=13.6~\textrm{as}$ for the $1$st pulse and $T_d=21.2~\textrm{as}$ for the $2$nd pulse.
Corresponding to the exponential spectrum region, the pulse spectral phase is manifested to be constant as shown in Fig.~\ref{normal_spectral_phase} (b).
At the emission instant, the sign of the incident electric field is $\hat{E}^{i}_{y}=-1$ for the $1$st pulse and $\hat{E^{i}_{y}}=1$ for the $2$nd pulse, and thus combining with Eq.~(\ref{Spectral_Phase_psi}), we can obtain:
\begin{subequations}
\label{simu_Spectral_psi_AM}
\begin{align}
\psi_{A_m}& \approx 0.35\pi\,,~~~~\textrm{for}~1\textrm{st pulse}\,,\\
\psi_{A_m}& \approx 0.30\pi\,,~~~~\textrm{for}~2\textrm{nd pulse}\,.
\end{align}
\end{subequations}
Making use of Eq.~(\ref{theo_AS_compact}) with the results in Eqs.~(\ref{simu_pulse_wd}) and (\ref{simu_Spectral_psi_AM}), and choosing the pulse amplitude $\bar{A}_m$ as
\begin{subequations}
\label{pulse_parameters_Am}
\begin{align}
\bar{A}_m &= 5.4\,,~~~~\textrm{for}~1\textrm{st pulse}\,,\\
\bar{A}_m &= 4.8\,,~~~~\textrm{for}~2\textrm{nd pulse}\,,
\end{align}
\end{subequations}
we completely reproduce the waveforms of $1$st and $2$nd attosecond pulses in Fig.~\ref{normal_pulse} (c) and (d) respectively.
Thus, we can verify the validity of our analytical model for the normal incidence case.

We try to estimate the pulse amplitude $\bar{A}_m$ from the simulation.
In Fig.~\ref{normal_pulse_wd} (a)-(d), we present the phase space distribution $f(\beta_x,\beta_y)$ and the density $n_e$ of the electron layer at the emission instants.
As shown, the phase space of the electron layer is expanded.
We assume that the emitted pulse is synthesized by the radiations from the electrons in a narrow space: $\beta_x<-0.8$ and $|\beta_y|<0.1$.
By integrating the electron density distribution [red and black lines in Fig.~\ref{normal_pulse_wd} (c) and (d)]: $n_{el}=\int n_e(x)dx$, we obtain: $n_{el}=6.58$ for the $1$st pulse and $n_{el}=7.75$ for the $2$nd pulse, and we calculate the Lorentz factor: $\gamma_{el}\approx (1-\beta^2_x)^{-1/2}=1.90$, $1.79$ for the $1$st and $2$nd pulses from the backward velocity $\beta_x^{m}$ in Fig.~\ref{Fig_comfirm} (d) at the emission instants (green stars). Thus we can obtain the pulse amplitude from Eq.~(\ref{theo_Am}): $A_m = 6.25$ for the $1$st pulse and $A_m =6.93$ for the $2$nd pulse, which qualitatively matches the values used in Eqs.~(\ref{pulse_parameters_Am}).
The difference may come from the overestimation of the electron phase space. Here, we ignore the temporal variation of the pulse amplitude.
As we can also see in Fig.~\ref{normal_pulse_wd} (e, red line) and (f, black line), the thickness of the electron layer is about $\Delta x\approx0.0025\lambda_l$ for the $1$st pulse and $\Delta x\approx0.0040\lambda_l$ for the $2$nd pulse, which correspond to the fluctuation thresholds: $\omega^{th}_{f} \approx 450\omega_l$ and $\omega^{th}_{f} \approx 260\omega_l$ for the $1$st and $2$nd pulses in Fig.~\ref{normal_spectral_phase}, satisfying Eq.~(\ref{threshold}) qualitatively.

\subsection{Oblique incidence}\label{oblique}
Now, we consider a more general situation where an ultraintense laser pulse with a long ramping front is obliquely ($\theta=45^{\circ}$) incident onto the pre-plasma, $n_{e}=n_c\exp{(x/L)}/2$, present in front of a plasma bulk, where $L$ is the scale length of the pre-plasma.
For convenience, we treat this interaction geometry in a Lorentz boosted frame~\cite{Bourdier:1983PoF} in which the plasma target moves with the initial velocity $\beta^{b}_y=-\sin(\theta)$ along the plasma surface and is irradiated normally by the laser pulse with p-polarization.

\begin{figure}
  \includegraphics[width=0.45\textwidth]{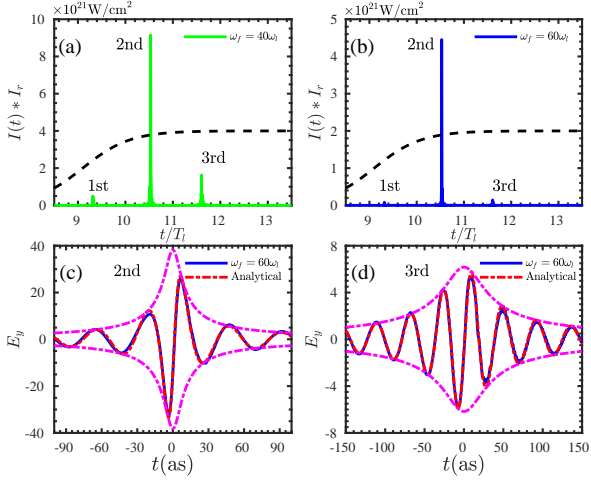}
 \caption{Attosecond pulses with different filter frequency: (a) $\omega_f=40\omega_l$, (b) $\omega_f=60\omega_l$. (c), (d) Electric field $E_y$ of the $2$nd and $3$rd pulses compared with the analytical expressions. We label the pulse center at time $t=0$ and zoom in the time axis in unit of as. The temporal envelope [see Eq.~(\ref{AS_profile}), magenta dashed-dotted line] of each pulse is also shown. The laser, $a(t)=a_{0}\sin(t+\phi)\{\tanh[(t-T_s)/W]-\tanh[(t-T_e)/W]\}/2$, radiates the plasma ($n_e=500n_c$, $L=\lambda_{l}/8$) with incident angle $\theta=45^{\circ}$, where $a_{0}=100$, $W=T_{l}=\lambda_{l}/c$, $T_s=6T_{l}$, $T_e=14T_{l}$, $\lambda_l=0.8\mu m$, and the CEP $\phi=0\pi$. The laser profile (black dashed line) is shown in (a) and (b) with a.u.. The field detector is located at $3\lambda_l$ from the plasma surface.}\label{oblique_pulse}
\end{figure}

\begin{figure}
  \includegraphics[width=0.48\textwidth]{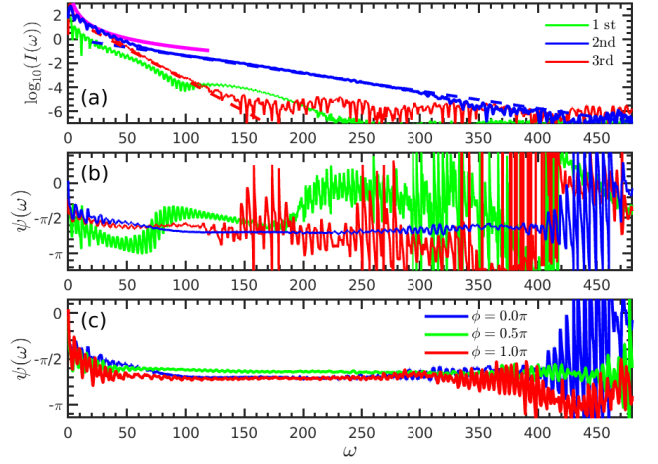}
 \caption{Intensity spectra (a) and spectral phase (b) for the pulses in Fig.~\ref{oblique_pulse}. The spectral fittings of the $2$nd pulse [$\textrm{log}_{10}(I)=0.1-0.0145\omega$, blue dashed line] and $3$nd pulse [$\textrm{log}_{10}(I)=1.9-0.052\omega$, red dashed line] are shown with the power-law spectral scaling [$I(\omega)\propto \omega^{-8/3}$, magenta solid line] fitting the lower-frequency region in the spectra. (c) Spectral phase of the $2$nd pulses in the cases driven by the lasers with different CEP $\phi$. The simulation in Fig.~\ref{oblique_pulse} is repeated with the same parameters, but different CEP ($\phi=\pi/2$, $\pi$).}\label{oblique_spec_phase}
\end{figure}

In Fig.~\ref{oblique_pulse}, we show the obtained attosecond pulses by filtering out low-order harmonics ($\omega < \omega_f$) in the reflected wave.
For the oblique incidence ($\theta=45^{\circ}$), only one attosecond pulse is emitted in each laser cycle because the laser electric field
in one half of the cycle hinders the compression of the electrons.
Here the $1$st pulse arises due to the reflection of the laser ramp which is too weak to compress a well-defined electron layer, thus leading to weak spectral intensity in Fig.~\ref{oblique_spec_phase} (a) and large phase fluctuations in Fig.~\ref{oblique_spec_phase} (b).
The $2$nd pulse is emitted in the first cycle of the main laser pulse interacting with the bulk of the plasma target, and the $3$rd pulse generated in the next laser cycle is emitted with much weaker intensity than the $2$nd pulse.
As seen in Fig.\ref{oblique_spec_phase} (a), the $2$nd pulse has a slower intensity decay than the $1$st and $3$rd pulses, implying higher efficiency for high-frequency emission.
In Fig.\ref{oblique_spec_phase} (b), the high-frequency components in the $1$st and $3$rd pulses display larger phase fluctuation than that in the $2$nd pulse, which could further reduce the pulse intensity and extend the duration. In the following laser cycles, the generated pulses would have much faster spectral decay and strong phase fluctuation~\cite{tang2018super}. 

\begin{figure}
\centering
  \includegraphics[width=0.45\textwidth]{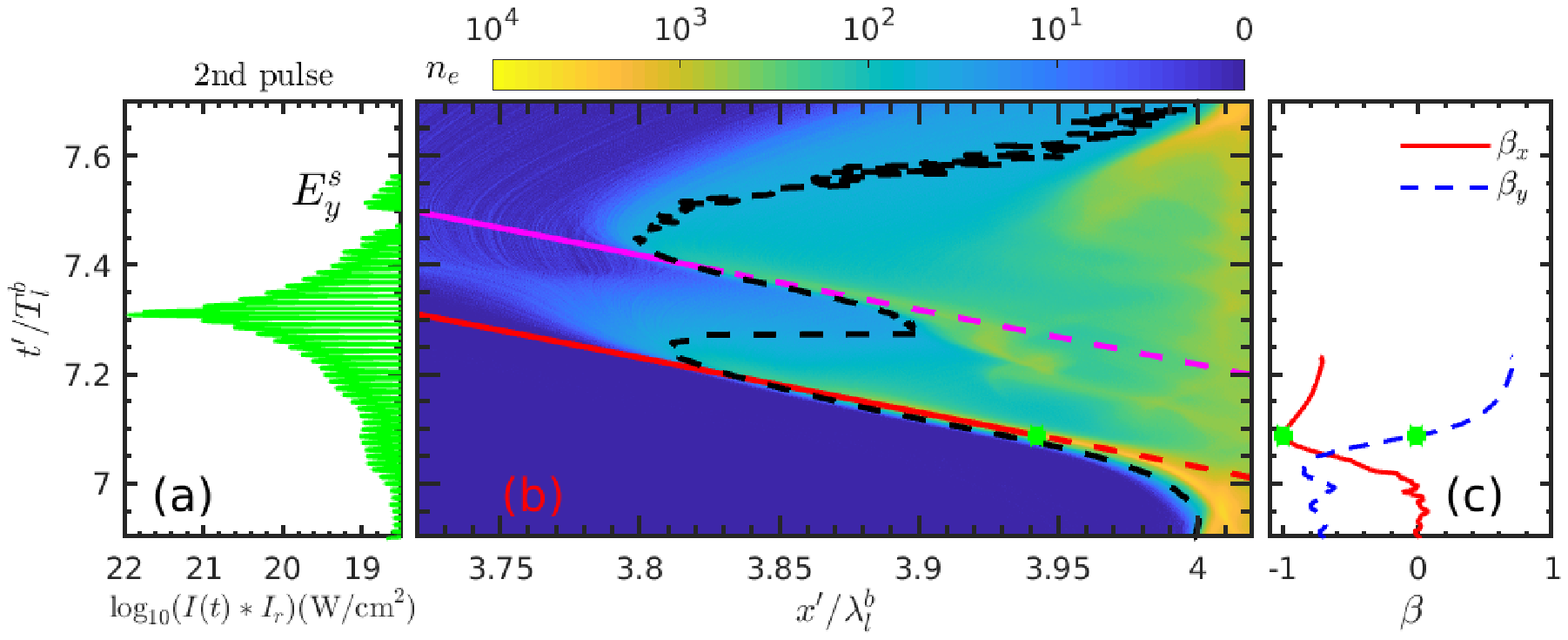}\\
  \includegraphics[width=0.45\textwidth]{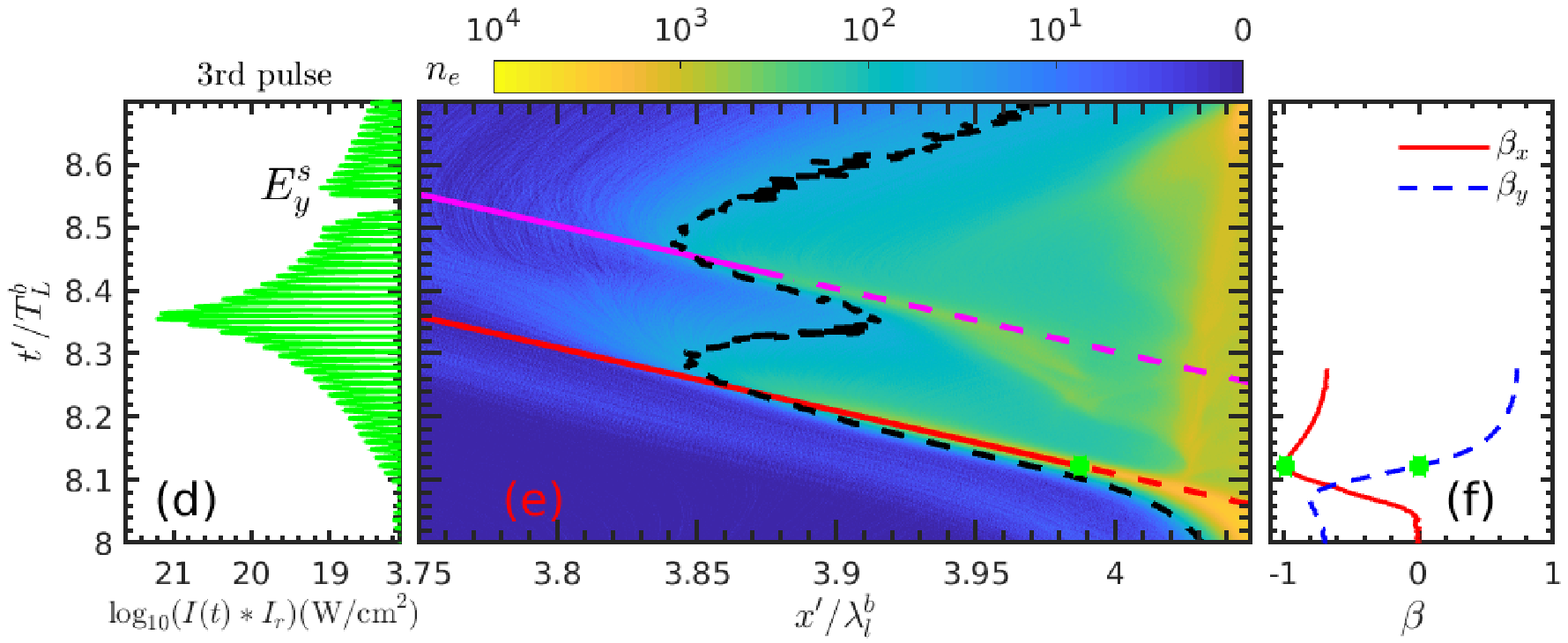}\\
  \includegraphics[width=0.45\textwidth]{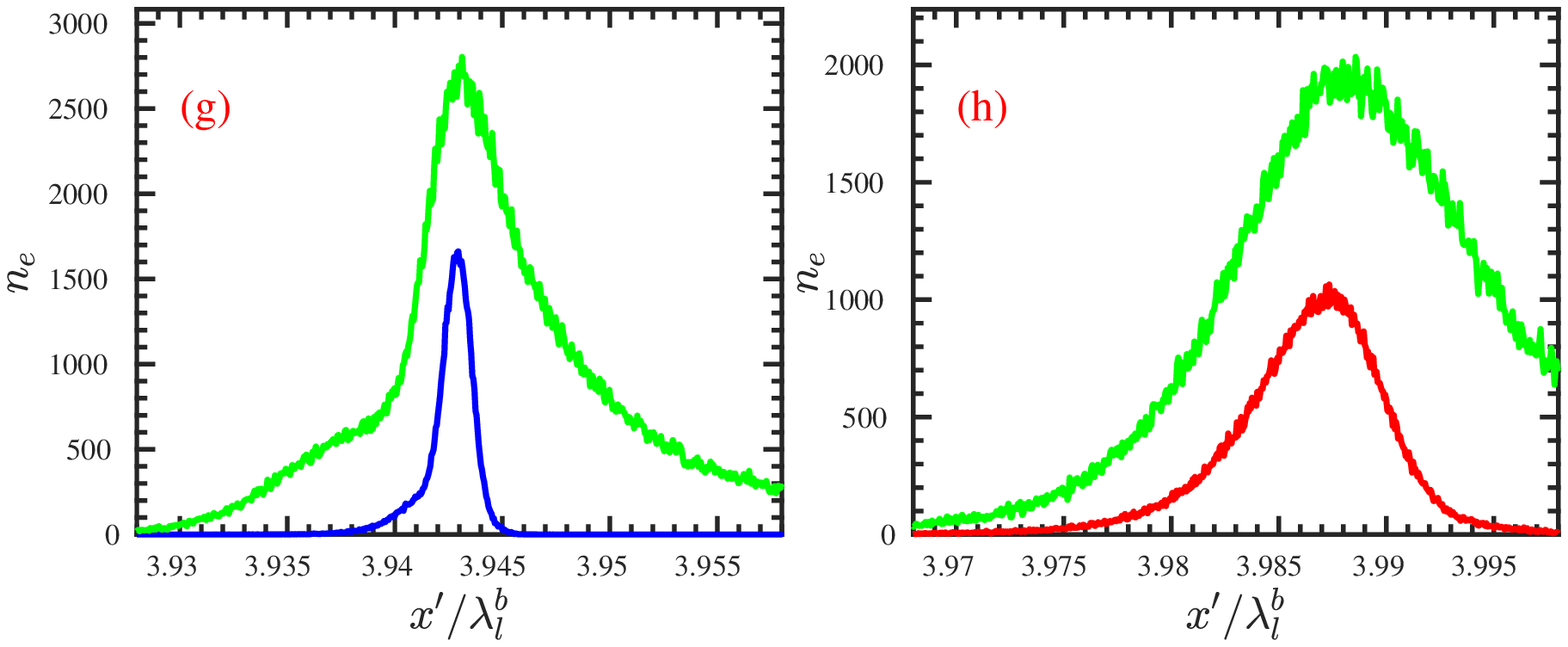}
 \caption{ $1$D PIC simulation for the $2$nd and $3$rd pulse emission processes in Fig.~\ref{oblique_pulse}.
 (a) Temporal shape of the $2$nd pulse in Fig.~\ref{oblique_pulse} (a) shifted along the retardation path [red line in (b)].
 (b) Evolution of the electron density $n_e$ at the plasma surface around the emission instant (green star) of the $2$nd pulse overlaid with the retardation paths of the pulse center (red line) and the subpulse [$E_y^s$ in (a), magenta line]. The evolution of the electron surface ($n_e = a_0 n^{b}_c$ ) is also shown (black dashed line).
 (c) Velocity ($\beta_x$, $\beta_y$) of the electron current along the retardation paths.
 (g) Density distribution of the plasma electrons at the emission instant (green line).
 In (d), (e), (f) and (h), we repeat all the plots in (a), (b), (c) and (g), but for the $3$rd pulse.
 In (c) and (f), we obtain $\beta_x^m=-0.9962$ and $\beta_x^m=-0.9904$ at the emission instants ($\beta_y=0$, green stars) of the $2$nd and $3$rd pulses.
 In (g) and (h), we also show the density of the electrons in the phase space: $\beta_x<-0.99$, $|\beta_y|<0.02$ for the $2$nd pulse (blue line) and $\beta_x<-0.97$, $|\beta_y|<0.06$ for the $3$rd pulse (red line).}\label{fig_model_oblique}
\end{figure}

In Fig.~\ref{fig_model_oblique}, we scan the emission processes for the $2$nd and $3$rd pulses in the boosted frame.
As shown, the pulses are clearly emitted by the surface strongly compressed electron layer when the layer transverse velocity $\beta_y$ changes sign and its backward velocity $\beta_x$ approaches the speed of light [see the green stars in Fig.~\ref{fig_model_oblique} (b), (c) for the $2$nd pulse, (e), (f) for the $3$rd pulse], which indeed satisfies the characters~[\ref{condition1},~\ref{condition2}] for our pulse emission model in Sec.~\ref{model_description}.
Since the contribution of net ion current to high frequencies is negligible due to its slow response to the electric field, all the derivations in Sec.~\ref{Theo_model} are repeatable in this boosted frame with the new normalization quantities: $k^{b}_l=k_l\cos(\theta)$, $\omega^{b}_l=\omega_l\cos(\theta)$, $n^{b}_c=n_c\cos^{2}(\theta)$.
Because of the Lorentz invariance of the normalized electric field ($eE/me\omega$) and time ($\omega t - k x$), Eq.~(\ref{theo_AS_compact}) can be applied directly for oblique incidence in lab reference:
\begin{align}
  E^{r}(\omega_f,t)=\frac{-2\hat{E^{i}_{y}}\bar{A}^{b}_m e^{-\frac{\omega_f}{\omega^{b}_d}}}{\sqrt{1+(\omega^{b}_dt)^2}}\cos\left[\omega_f t+\varphi(t)-\psi_{A_m}\right]
\label{AS_pulse_oblique}
\end{align}
but with $\bar{A}^{b}_m$ and $\omega^{b}_d$ calculated in the boost frame by Eqs.~(\ref{theo_Am}) and (\ref{theo_wd}).

In Fig.~\ref{oblique_spec_phase} (a), the pulse exponential spectra are confirmed by the linear-logarithm fittings in the regions: $50<\omega<400$ for the $2$nd pulse (blue dashed line) and $40<\omega<110$ for the $3$rd pulse (red dashed line), and in Fig.~\ref{oblique_spec_phase} (b) the pulse spectral phase in the corresponding regions are proved to be constant.
The fitting slopes of the pulse spectra give the spectral decay:
\begin{subequations}
\label{oblique_pulse_wd}
\begin{align}
\omega^{b}_d &\approx 58.82\,,~~~~\textrm{for}~2\textrm{nd pulse}\,,\\
\omega^{b}_d &\approx 16.66\,,~~~~\textrm{for}~3\textrm{rd pulse}\,,
\end{align}
\end{subequations}
which fix the pulse temporal envelopes in Fig.~\ref{oblique_pulse} (c) and (d), and reveal the duration $T_d=2/\omega^{b}_d=14.2~\textrm{as}$ for the $2$nd pulse and $T_d=50.8~\textrm{as}$ for the $3$rd pulse.
At the emission instants, the sign of the incident electric field is $\hat{E}^{i}_{y}=-1$ for each pulse. With Eq.~(\ref{Spectral_Phase_psi}) and Fig.~\ref{oblique_spec_phase} (b), we quantitatively obtain that:
\begin{subequations}
\label{oblique_psi_AM}
\begin{align}
\psi_{A_m}&\approx 0.12\pi~~~~\textrm{for}~2\textrm{nd pulse}\,,\\
\psi_{A_m}&\approx 0.06\pi~~~~\textrm{for}~3\textrm{rd pulse}\,.
\end{align}
\end{subequations}
Inserting Eqs.~(\ref{oblique_pulse_wd}) and (\ref{oblique_psi_AM}) into Eq.~(\ref{AS_pulse_oblique}), and choosing
\begin{subequations}
\label{Eq_pulse_Am_oblique}
\begin{align}
\bar{A}^{b}_m &= ~52~~~~\textrm{for}~2\textrm{nd pulse}\,,\\
\bar{A}^{b}_m &= 112~~~~\textrm{for}~3\textrm{rd pulse}\,,
\end{align}
\end{subequations}
we validate our pulse emission model for oblique incidence case by precisely reproducing the $2$nd and $3$rd attosecond pulses in Fig.~\ref{oblique_pulse} (c), (d).

To access the pulse amplitude $A^{b}_m$ from the simulation, we gather the electrons in the phase space: $\beta_x<-0.99$, $|\beta_y|<0.02$ for the $2$nd pulse and $\beta_x<-0.97$, $|\beta_y|<0.06$ for the $3$rd pulse, and plot the density distributions $n_e$ respectively in Fig.~\ref{fig_model_oblique} (g, blue line) and (h, red line).
By integrating the density distributions, we gain: $n_{el}=13.5$, $36.7$ for the $2$nd and $3$rd pulses, and with the backward velocity $\beta^{m}_x$ in Fig.~\ref{fig_model_oblique} (c) and (f) at the emission instants (green stars), we calculate the Lorentz factor: $\gamma_{el}\approx 11.4$, $7.2$ for the $2$nd and $3$rd pulses.
Based on Eq.~(\ref{theo_Am}), we can obtain: $A^{b}_m \approx 77$ for the $2$nd pulse and $A^b_m \approx 132.1$ for the $3$rd pulse, which are close to the values used in Eqs.~(\ref{Eq_pulse_Am_oblique}).
We can also see that the thickness of the electron layer is about $\Delta x\approx0.002\lambda^{b}_l$ for the $2$nd pulse and $\Delta x\approx0.007\lambda^{b}_l$ for the $3$rd pulse, which correspond to the phase fluctuation thresholds $\omega^{th}_{f} \approx 400\omega_l$ and $\omega^{th}_{f} \approx 150\omega_l$ for the $2$nd and $3$rd pulses as shown in Fig.~\ref{oblique_spec_phase} (b).

\subsection{Discussion}
In Fig.~\ref{normal_pulse_wd} (e) and (f), we compute the value of $\omega_d$ based on the cold fluid approximation in Eq.~(\ref{theo_wd_cold}) along the pulse retardation paths in Fig.~\ref{Fig_comfirm}.
At the emission instants (green stars), the obtained values qualitatively match the values in Eqs.~(\ref{simu_pulse_wd}), but are larger.
This may be because the plasma thermal pressure and collision friction decreases the transverse acceleration of the electron layer [see Eq.~(\ref{motion_1d})].
In the ultra-relativistic case in Fig.~\ref{fig_model_oblique}, these two terms become more important as the calculations of the parameter $\omega_d$ in Eq.~(\ref{theo_wd_cold}) (not shown) are much larger than the values obtained from the pulse spectra in Eq.~(\ref{oblique_pulse_wd}).
Here, we have to state that the accurate specification of the emission instant is difficult, which may also induce numerical error in the calculation of $\omega_d$.

As we see in Fig.~\ref{normal_spectral_phase} (a) and Fig.~\ref{oblique_spec_phase} (a), the pulse intensity spectra are larger than the extension of the fitting spectra (dashed lines) outside the exponential region.
In the lower-frequency region, this is because more low frequencies are radiated by the electron layer during its backward acceleration, leading to the power-law spectral scaling $I(\omega)\propto \omega^{-8/3}$ (magenta solid lines)~\cite{Baeva:2006PRE}.
In Fig.~\ref{normal_spectral_phase} (b) and Fig.~\ref{oblique_spec_phase} (b), phase-chirp~\cite{Varju:2005JMO} is manifested in the corresponding frequency region resulting from the different Doppler frequency upshift at different times during the acceleration.
In higher-frequency region, the more intense spectra come from the superposition of the radiations from all the electrons in the surface layer [green lines in Fig~\ref{normal_pulse_wd} (a) (b) and Fig.~\ref{fig_model_oblique} (g) (h)].

As discussed in Sec.~\ref{Theo_spectral_Phase} and validated with the simulation results in Fig.~\ref{normal_spectral_phase} (b) and Fig.~\ref{oblique_spec_phase} (b), the emitted pulse has a constant spectral phase in a rather broad frequency region.
The value of the constant phase $\psi(\omega)=\pm\pi/2-\psi_{A_m}$ is determined by the dynamics of the well-defined electron layer and slightly perturbed by the temporal variation of pulse amplitude $A_m(t)$.
In Fig.~\ref{oblique_spec_phase} (c), we show the spectral phase of the emitted pulses in the cases driven by the lasers with different CEP.
As we can see, with the laser CEP changing from $\phi=0.0\pi$ to $\phi=1.0\pi$, the value of the spectral phase in the region $50\lesssim \omega \lesssim 400$ displays a small change less than $\Delta \psi \approx 0.1\pi$.
This indicates that with an appropriate frequency filter $\omega_f > 50$, the obtained attosecond pulse has a quasi-stable constant spectral phase independent on the laser CEP, which highlights the applications of the emitted attosecond pulse in experiments avoiding shot-to-shot changes due to the unstable laser CEP~\cite{heissler:2010APB,*Hoff:2017Natph}.
Moreover, we can also see that the coherent spectral interval can be extended to a higher-frequency region ($\omega^{th}_{in} \approx 450$) with a well controlled laser CEP ($\phi=\pi/2$).

In Fig.~\ref{fig_model_oblique} (a) and (d), we clearly see the double-pulse structure of the emitted pulse: a subpulse $E_y^s$ is emitted after the main pulse by the secondary electron bunch formed behind the electron layer~\cite{Cousens:2016PRL} as shown by the intersection between the retardation path of the subpulse (magenta line) and the electron surface (back dashed line) in Fig.~\ref{fig_model_oblique} (b) and (e).
The interference between the double pulses in same emission process would lead to the oscillations in the pulse spectra and spectral phase.
Due to the less efficiency of the secondary bunch radiation~\cite{tang2018super}, these oscillations can only be present in the low-frequency region as shown in Fig.~\ref{oblique_spec_phase}.
To get an attosecond pulse with a constant phase, a suitable frequency filter is needed to overcome the phase oscillation.
We wish to stress that these oscillations are the consequence of the interference between the double pulses, which is essentially different from the plasma-wave modulation in the harmonic spectrum of the total reflection~\cite{Boyd:2000PRL,*Watts:2002PRL,*Boyd:2007PRL}.

\section{Summary}
We develop an analytical model which describes the attosecond pulse emission from a strongly compressed electron layer in the highly relativistic laser-plasma interaction. The physical properties of the emitted attosecond pulse are completely described: exponential spectrum, constant spectral phase and explicit waveform [see Eqs.~(\ref{theo_spectrum}), (\ref{Spectral_Phase_psi}) and (\ref{theo_AS_compact}) respectively]. All of these properties are validated by PIC simulations for both normal and oblique incidence cases.

From the simulation results, we clearly see that the emitted attosecond pulse possesses a phase-stabilized spectrum ranging from $\omega \approx 40 \sim 60$eV to $\omega \approx 400 \sim 600$eV and highly relativistic intensity $I \gg I_r$, which could promote the attosecond metrology to the ultrafast physical processes in x-ray regime. We also see that, with an appropriate frequency filter $\omega_f$ [see Fig~\ref{oblique_pulse} (b)], the attosecond pulse emitted in the first cycle of the main laser pulse is order of magnitude stronger than those emitted in the following cycles, which highlights the potential to generate an isolated ultraintense attosecond pulse. The choice of the frequency filter for the pulse isolation is conducted by the parameter maps in Ref~\cite{Tang:2017PRE}.

%
\end{document}